\documentclass[a4paper, 11pt]{article}    
\usepackage{latexsym}
\usepackage{amssymb}
\usepackage{amsmath}
\usepackage{amsfonts}
\usepackage{graphicx}
\usepackage{float}
\usepackage[english]{babel}

\newcommand{\quota}[1]{``#1''}

\title{ \bf Restructuring the \quota{one-way CSA} counterparty risk in a CDO}  
\author{ Lorenzo Giada \\ Banco Popolare, Verona \\ {\tt lorenzo.giada@gmail.com}$^*$ 
\and Claudio Nordio \\ Banco Popolare, Verona \\ {\tt c.nordio@gmail.com}\footnote{This paper reflects the authors' opinions and
not necessarily those of their employers.}}

\begin{document}           
\maketitle                 

\begin{center}
\tt \Large Working paper
\end{center}

\vspace{10mm}

\begin{abstract}
\it \noindent
We show how to restructure the counterparty risk faced by the originator of a securitization or covered bond arising from an interest rate hedging swap assisted by a \quota{one-way} collateral agreement. This risk emerges when the swap is negotiated between the special purpose vehicle and a third party that covers itself through a back-to-back swap with the originator. We show that the counterparty risk of the originator may be removed by adding a chain of back-to-back credit derivatives between the three parties (originator, counterparty and vehicle).
\end{abstract}

\bigskip

{\bf JEL} Classification codes: G13, G18, G33 

{\bf AMS} Classification codes: 91B99

\bigskip


\vspace{15mm}

\section{Summary}
As a Special Purpose Vehicle (SPV) issues a securitization, it has to enter into a hedging interest rate swap (a basis swap) that transforms the interest payments coming from the collateral assets into the coupons of the issued note. When such a derivative is traded with a third party (different from the original owner of the collateral assets, i.e. the originator), the latter usually enters into a back-to-back swap with the originator, and both trades are supported by \quota{one way} CSAs that, on one hand do not require cash collateral from the SPV, but on the other hand do not protect the originator from the default of its counterparty \cite{onewaycsa3}. The counterparty risk faced by the originator can therefore be quite large, taking into account the typically large notional of such deals and their long maturity. All of the above also apply to the case of a covered bond, and it is well known that the long term transactions assisted by asymmetric collateralization agreements imply a considerable impact in term of capital charge for the unhedged party, especially under the new Basel III framework coming in force \cite{onewaycsa1}, \cite{onewaycsa2}. In the last years, the European financial institutions have made large use of securitizations and covered bond to improve their liquidity, and the deterioration of the credit quality of many originators forced the recourse to third parties with better rating as swap counterparties, in order to mantain high rating grades for the SPVs.\\
\newline
In the following we propose a stylized securitization structure that eliminates such counterparty risk even if it still does not require any collateral to be posted to the originator. With minor changes it is applicable also to the case of a covered bond. It consists in a set of credit derivatives transactions replicating a three party agreement (TPA) between originator, SPV and swap counterparty that is fair for all parties and that prevents any liquidity shock that may arise following the default of the swap counterparty. Even if the TPA itself would represent a financial hedge of the counterparty risk, the set of credit derivatives seems to us a financial equivalent but more robust structuring under the bankruptcy law. We will show that it represents an effective restructuring of the one-way counterparty risk linked to the securitization/covered bond structure, therefore permitting a considerable capital relief.

\section{Swap's counterparty default in the usual structure}
We examine the case in which a Counterparty (C) enters into a hedging  swap (the Front Swap) with the SPV (V), the issuer of the Note, and hedges its position with a back-to-back swap (the Back Swap) with the Originator (O). Both trades are supported by independent ISDA agreements with \quota{one-way} CSA.

\begin{figure}[h]
\centering
\includegraphics[width=0.50\textwidth]{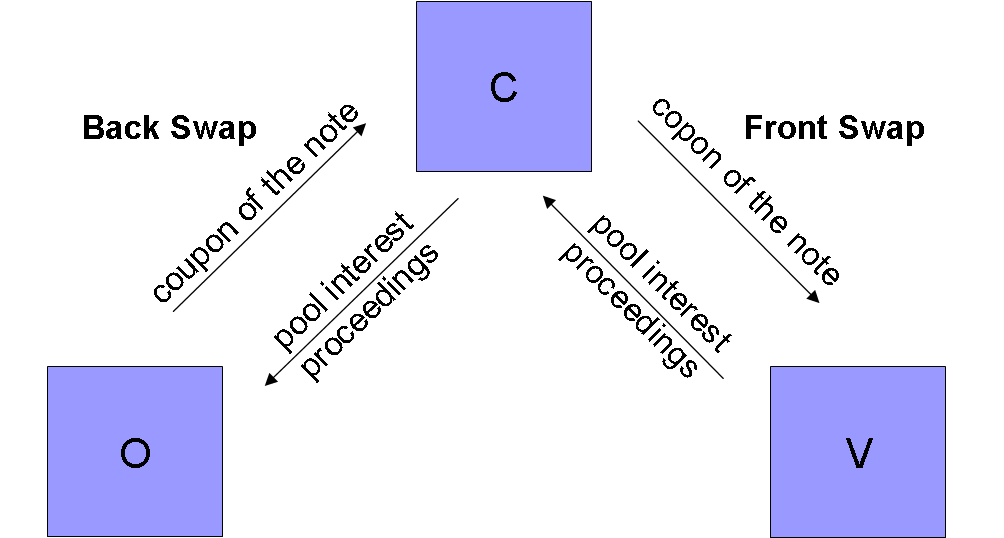}
\caption{The swap hedging the interest rate risk of the SPV (Front Swap) and its back-to-back transaction (Back Swap).\label{fig1}}
\end{figure}

Neglecting the fees paid running to C by O, we describe a stylized version of the structure. In the Front Swap V pays to C the interest proceedings from the securitized assets, and C pays to V the coupons of the Note (see Figure \ref{fig1}). Correspondingly, in the Back Swap C pays back to O the interest proceedings, and receives the coupons of the Note. These two contract are set up only in order to hedge the interest rate risk of the SPV, that otherwise would face a potential liquidity mismatch. To guarantee the patrimonial segregation of the SPV, the ISDA agreement and CSA between C and O and those between C and V are distinct, and the CSAs have a threshold structure that is  constructed in such a way as to prevent V from posting any collateral to C and correspondingly to prevent C from posting collateral to O (the \quota{one-way} CSA mentioned above). The reason for this is that, typically, the SPV has a small liquidity buffer that might be insufficient to post the required collateral to C, but the high creditworthiness of the former makes the risk acceptable for the latter, even in absence of collateralization. However, in order to minimize the cost of the transaction to O, C will not be required to post collateral to O either. For the sake of simplicity we will assume henceforth that the CSAs guarantee a perfect counterparty risk\footnote{ i.e. no gap risk, and continuous collateralization with zero threshold and minimum transfer amount} hedge in case the collateral has to be posted by O to C and by C to V.\\
\newline
Rating agencies require a high credit standing for the swap counterparty C, but its default probability can be significantly different from zero. This generates counterparty risk for O, taking into account the fact that the Back Swap can take large mark-to-market values, given the large notionals that are typical for this kind of transactions. The consequences of this affect both capital requirements through the risk weighted asset (RWA) related to the exposure to the counterparty C, and the P\&L statement through the expected loss (CVA) related to such exposure. Moreover, the Basel III framework has introduced additional capital requirements related to the variability of this component, the so-called CVA risk.\\
\newline
In the following we will examine (in a stylized way) the economic consequences of the default of C both when the mark-to-market of the Back Swap is negative for O, and when it is positive. In the first case, assuming as before perfect hedge of counterparty risk via the CSA, the collateral posted by O to C and by C to V equals the mark-to-market of the Front Swap as seen by V, which amounts to the opposite of the mark-to-market of the Back Swap as seen by O. Therefore, the {\it Termination Amounts} as defined in the ISDA agreement for the Originator and the SPV are both zero. Moreover, if we assume that the replacement transactions are put in place instantaneously, neither O nor V will experience a liquidity shock, since on one hand O will replace the liquidity posted to C with the upfront of the {\it Replacement Transaction}, and on the other hand V will pay as upfront the collateral posted by C to the new counterparty (see Figure \ref{fig3}).

\begin{figure}[h]
\centering
\includegraphics[width=0.50\textwidth]{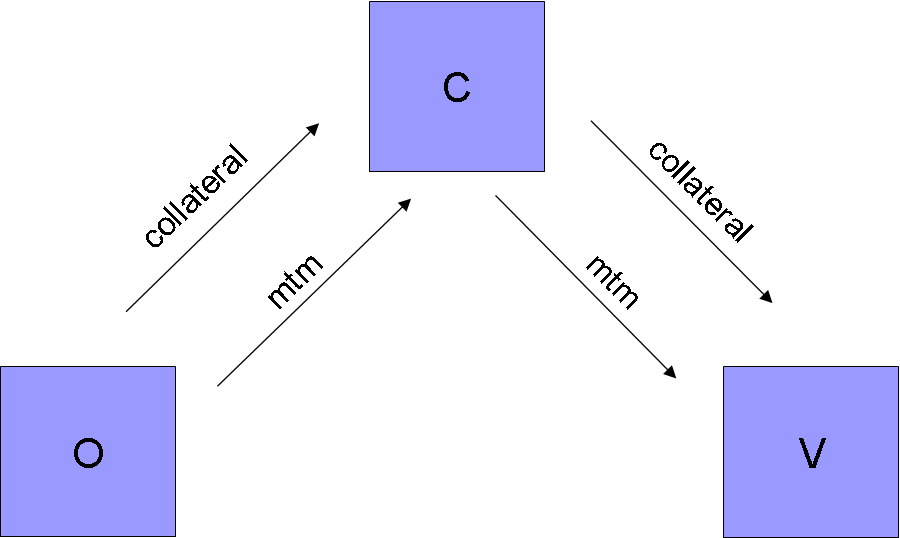}
\caption{The one-way mechanism. Each arrow indicates the party for which the mark-to-market is positive and to which the collateral is posted. Here the mark-to-market of the Back Swap is negative for O, and the collateral is posted as usual. Under the assumption of perfect collateralization, no losses are suffered by the parties if C defaults.\label{fig3}}
\end{figure}

In case of default of C when the mark-to-market of the Back Swap is positive for O, the latter suffers a loss that amounts to a fraction (so-called \quota{loss given default}, LGD) of the mark-to-market, while the SPV does not experience any loss, nor any liquidity shock, since the {\it Termination Amount} now is paid by V to C, but is received by V from the new counterparty as the upfront of the {\it Replacement Transaction}.

\begin{figure}[h]
\centering
\includegraphics[width=0.50\textwidth]{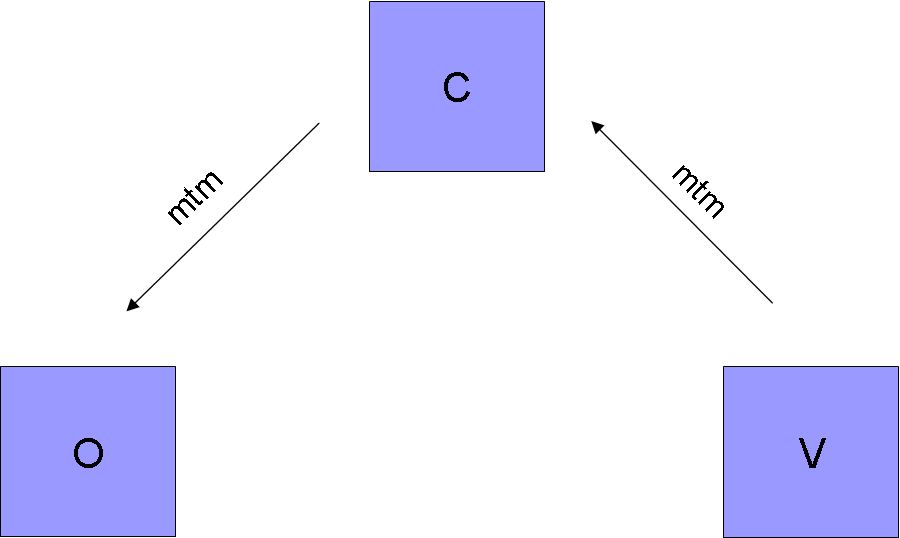}
\caption{The one-way mechanism. Each arrow indicates the party for which the mark-to-market is positive. Here the mark-to-market of the Back Swap is positive and no collateral is posted. If C defaults, V terminates the swap without losses, but O suffers a loss of LGD$\cdot$mtm.\label{fig3}}
\end{figure}

\section{A TPA as a hedge of the counterparty risk}
As mentioned before, even if the SPV and the Originator belong to the same financial group, there cannot be netting of the amount due by C to O with that due by V to C in case of default of C, since the two transactions refer to different ISDA agreements. In the following we will describe\footnote{an agreement containing clauses similar to the subsequent 1. and 2. was in place between Lehman Brothers SpA and a primary Italian commercial bank, and terminated out-of-money after the default of the former} an hypothetical three-party-agreement (TPA) that will set the basis for such a netting, and that introduces an additional ingredient required to ensure fairness for all parties.\\
\newline
The TPA consists of three clauses, that would be activated {\it only in case of default of C and if the mark-to-market of the Back Swap is positive for O}. If such event takes place:

\begin{enumerate}
\item O waives such amount in favour of C when it calculates the Termination Amount of the Back Swap;
\item C waives such amounts in favour of V when it calculates the Termination Amount of the Front Swap;
\item V pays to O the amount received as upfront from its new counterparty in the Replacement Transaction.
\end{enumerate}

It is easy to realize that the TPA constitutes a perfect hedge of the counterparty risk for O in the event that the mark-to-market of the Back Swap is positive, and at the same time it is fair for each party, and prevents any liquidity shock, in particular for the SPV. Let us point out that clause (A) together with the ISDA between C and O corresponds to the ISDA 1992 First Method, and the same is true for clause (B) and the agreement between C and V. However, this equivalence is no more true when both clauses are considered together. In fact they effectively implement the netting of the amounts owed by V and due to O. Moreover they are fair for C, that does not realize any profit or loss from its own default. Clause (C) is required to prevent the SPV from realizing a profit, and O from posting a loss, and it constitutes a form of netting between O and V. However it does not affect the solvency of the SPV nor its credit standing.

\section{The TPA as a chain of Contingent CDS}
Notwithstanding the TPA examined above represents a fair financial hedge to the counterparty risk of the Originator, we observe that its contractual design might not be considered robust under some bankruptcy law’s principle or interpretation. Indeed the bankruptcy supervisor of the Counterparty, when addressing the obligation related to the TPA after the default of the Counterparty, might identify an improper advantage for the Originator with respect to the other creditors, considering that such advantage is not linked to a payoff of an agreement rather than, more properly, to a deed of waiver. However, in what follows we illustrate an equivalent structuring realized through standard ISDA credit derivatives, which therefore may be preferrable to the TPA as more legally robust.\\
\newline
By entering in the usual structure without the TPA introduced above, the counterparty risk faced by the Originator is equivalent to a short position on an upfront Contingent Credit Default Swap (CCDS), i.e. a CDS whose notional is the mark-to-market (if positive) of a reference swap at the time of default of a reference entity, and whose premium has to be payed upfront. More specifically, the Originator gives protection to the Counterparty through a CCDS referenced to the Back Swap and to the Counterparty itself. Incidentally, in the usual structure showed above, the Counterparty gives protection to the SPV through a CCDS referenced to the Front Swap and to the SPV itself, but we may consider the CCDS be worth zero i.e. the SPV non defaultable, and hereafter we focus instead on the defaultability of the Counterparty.

\begin{figure}[h]
\centering
\includegraphics[width=0.50\textwidth]{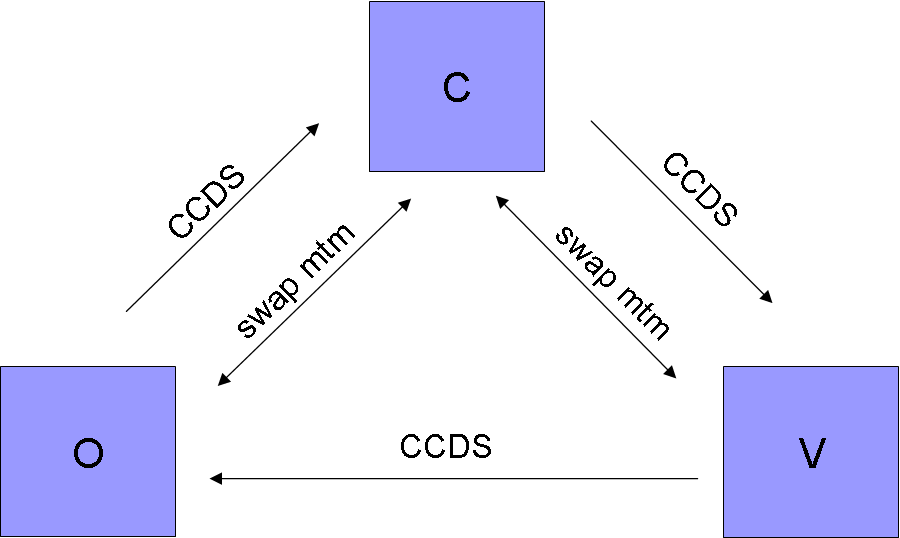}
\caption{The chain of contingent credit default swaps (CCDS) replicating the TPA. The CCDSs have the same terms and are referenced to C and the Back Swap; each related arrow indicates the protection buyer\label{fig2}}
\end{figure}

By entering the TPA, the Originator buys back through the CCDS the protection on the Counterparty’s default from the SPV, which buys it from the Counterparty itself, which hedges itself with the Originator. In other words, the three parts of the TPA listed above map into the following set of transactions involving identical CCDS referenced to the Back Swap and the Counterparty (see Figure \ref{fig2}):

\begin{enumerate}
\item C buys protection on itself from O through the CCDS;
\item V buys protection on C from C through the CCDS;
\item O buys protection on C from V through the CCDS.
\end{enumerate}

Note that all we said above about the TPA remains true in the present setting, and none of the parties pays any net premium, all being hedged through back-to-back transactions. Note however that in 2.~the SPV buys protection on the Counterparty from the Counterparty itself, but this issue can be addressed by negotiating the CCDS under the same ISDA agreement to which the Front Swap is referenced, so that the Front Swap itself fully collateralizes the credit default swap. Indeed, the ISDA agreement is a netting agreement allowing - in case of default of the Counterparty - the offset between the payoff of the credit default swap\footnote{such payoff, triggered by the default of the Counterparty, is defined as the positive part of the reference swap and, given that its reference name is the Counterparty itself, in the case of transaction 2 it has to be considered as a Unpaid Amount under the ISDA 1992 definitions} and the mark-to-market of the Front Swap. The former is negative for the SPV when the Back Swap’s one is positive and therefore equal to the payoff of the CCDS; conversely, if the Counterparty defaults when the mark-to-market of the Front Swap is positive for the SPV, the CCDS expires worthless and, as a consequence of the perfect collateralization we assumed before, the SPV does not suffer any loss.

\begin{figure}[h]
\centering
\includegraphics[width=0.50\textwidth]{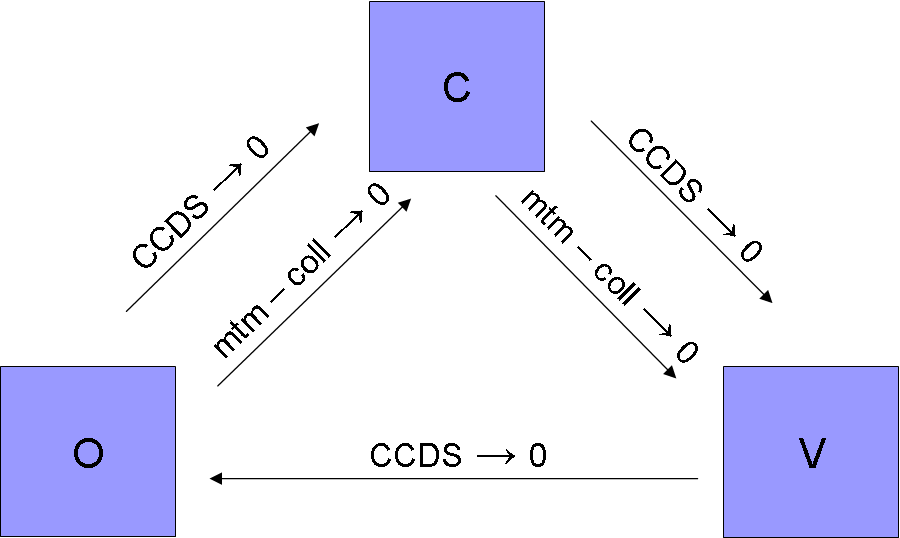}
\caption{The termination amounts, in case of negative mark-to-market of the Back Swap for O.\label{fig2}}
\end{figure}

Analogously, also the CCDS in 1.~must be negotiated under the same ISDA agreement of the Back Swap, and if its mark-to-market is positive in case of default of the Counterparty it collateralizes the payoff of the CCDS, protecting the Counterparty. This resembles the case of a counterparty buying a CDS from and referenced to another counterparty, collateralized by an amount equal to the entire notional posted by the latter to the former: for a contingent CDS, at the time of default the notional is indeed the positive part of the reference swap.\\
\newline
The ideas illustrated so far are ultimately based of the observation that, if a party enters in a swap with another party, the \quota{unfunded} counterpart of a \quota{funded} asymmetric hedge given by a \quota{one-way} collateral agreement between such parties, is just a CCDS exchanged between the parties and referenced to one of them, under the same netting agreement with the reference swap. In other words the following two may be thought as equivalent (see also \cite{fbva}):
\begin{itemize}
\item A enters in a swap with B, with a \quota{one-way} collateral such that a perfect collateral is posted from A to B, therefore B is fully hedged by counterparty risk but A remains unhedged;
\item A enters in a swap with B, and B buys - from A under the same netting agreement - a CCDS referenced to the swap and to A.
\end{itemize}
This example should be considered in analogy with the transaction 2.~above (the SPV as A, the Counterparty as B). A similar argument may be outlined for the transaction 1. We may say that, instead of restructuring the preexistent \quota{one-way} CSA into an ordinary \quota{two-way} CSA by adding a second \quota{one-way} CSA on the opposite direction as in the first bullet point above (not a viable solution, given the liquidity squeeze of the SPV), our proposal makes use of its \quota{unfunded} counterpart. Finally, a back-to-back strategy is made possible given the peculiar three-party relationship between the originator, the counterparty and the vehicle.

\begin{figure}[h]
\centering
\includegraphics[width=0.50\textwidth]{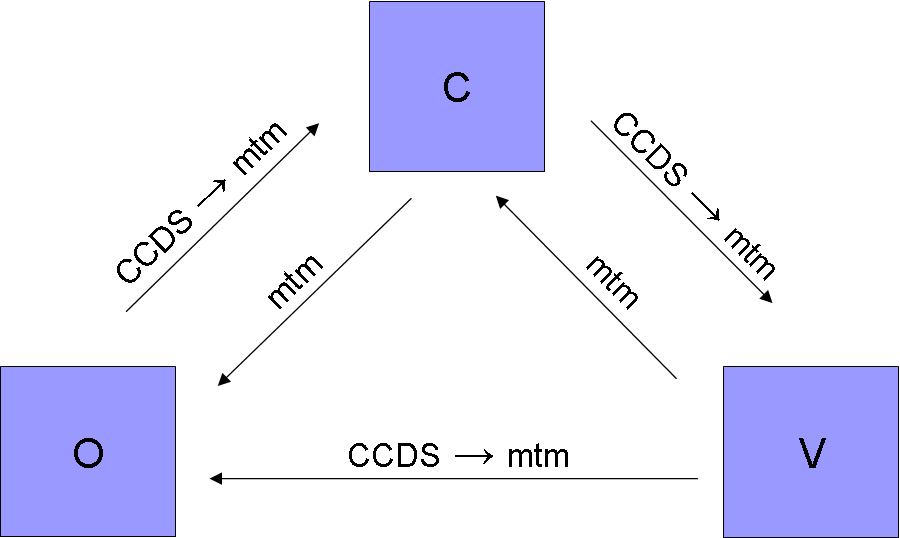}
\caption{The termination amounts, in case of positive mark-to-market of the Back Swap for O.\label{fig2}}
\end{figure}

\section{Conclusions}
When a special purpose vehicle enters in a swap with a counterparty in order to hedge the interest rate risk related to the issue of a securitization or a covered bond, usually such counterparty enters in turn in a back-to-back transaction with the originator, generally assisted by a \quota{one-way} collateral, leaving the originator to face a generally remarkable counterparty risk. We showed how this risk may be canceled, without paying any premium, by entering in a chain of contingent credit default swaps (CCDS), at the same time bought and sold by each party, moving from the basic idea that the effects on the default risk of a counterparty of an asymmetric collateralized swap may be replicated by a CCDS referenced to and negotiated with the same counterparty, under the same netting agreement of the swap itself.

\section*{Acknowledgments}
We are grateful to Damiano Brigo for helpful discussions and Carlo Palego for encouraging our research.

\end{document}